\documentclass[proof]{WileyASNA-v1}

\usepackage{amsmath}	% Advanced maths commands

\articletype{Original Paper}%

\received{15 May 2025}
\revised{2 June 2025}
\accepted{1 July 2025}

\begin{document}

\title{High resolution radio analysis of the starburst galaxy NGC~4527: signatures of an AGN core}

\author[1,2]{Galante, C. A.}

\author[1,2]{Romero, G. E.}

\author[2]{Saponara, J.}

\author[2]{Benaglia, P.}

\authormark{Galante \textsc{et al.}}

\address[1]{\orgdiv{Facultad de Ciencias Astronómicas y Geofísicas}, \orgname{Universidad Nacional de La Plata}, \orgaddress{\state{Buenos Aires}, \country{Argentina}}}

\address[2]{\orgdiv{Instituto Argentino de Radioastronomía}, \orgname{CONICET-CIC-UNLP}, \orgaddress{\state{Villa Elisa, Buenos Aires}, \country{Argentina}}}

\corres{C. A. Galante. \email{cgalante@iar.unlp.edu.ar}}

\abstract[ABSTRACT]{NGC~4527 is a nearby edge-on spiral galaxy with both starburst and AGN features, hosting a LINER nucleus. We present a radio study of the large-scale structure and nuclear region of this galaxy, based on new uGMRT observations at 700 and 1230~MHz. Our continuum maps reveal extended emission tracing the stellar disk, with no evidence of a radio halo. The spectral index distribution and the presence of PAHs across the disk are consistent with ongoing star formation. In the nuclear region we resolve three compact sources: one at the galactic centre and two symmetrically aligned with the major axis at $\sim400$~pc. The spectral index values and the destruction of PAHs in the central source, together with previously detected X-rays emission, suggest the presence of a low-luminosity AGN. The two off-centre sources are consistent with a star formation ring, coincident with a molecular gas ring previously reported. We explore a scenario where super-Eddington accretion onto the black hole drives a dense wind that falls back onto the disk, triggering star formation in a circumnuclear ring.}

\keywords{Galaxies: individual (NGC~4527) -- galaxies: nuclei -- galaxies: starburst -- radio continuum: galaxies}

%\jnlcitation{\cname{%
%\author{Galante, C. A.}, 
%\author{Romero, G. E.}, 
%\author{Saponara, J.}, and 
%\author{Benaglia, P.}} (\cyear{2025}), 
%\ctitle{High resolution radio analysis of the starburst galaxy NGC~4527: signatures of an AGN core}. \cjournal{Astronomische Nachrichten}.}
%\ctitle{High resolution radio analysis of the starburst galaxy NGC~4527: signatures of an AGN core?}, \cjournal{Astronomische Nachrichten}, \cvol{}.}

\maketitle

\footnotetext{\textbf{Abbreviations:} AGN, active galactic nucleus; LINER, low-ionization nuclear emission-line region; uGMRT, upgraded Giant Metrewave Radio Telescope; MHz, megahertz; PAH, polycyclic aromatic hydrocarbons}

\section{Introduction}\label{sec:intro}

The interplay between active galactic nuclei (AGNs), powered by accretion onto supermassive black holes (SMBHs), and nuclear starbursts, fuelled by collapsing molecular clouds, shapes the dynamics of galactic centres \citep{Farrah2003,Scoville2003,Ishibashi2016,Gowardhan2018,Ishibashi2022}. AGNs inject energy into their surroundings through jets, winds, and radiation, while starbursts drive winds and enrich the interstellar medium (ISM) with supernova ejecta. In systems where these processes coexist, they compete for gas reservoirs, either suppressing or triggering star formation, and producing overlapping observational signatures across the electromagnetic spectrum \citep{Torbaniuk2021,Magliocchetti2022,Mountrichas2024,Nandi2025}. 

Similarly, gravitational instabilities in gas-rich nuclei -- amplified by bars or tidal interactions -- can channel gas inflows that simultaneously fuel starbursts and obscure AGNs, leading to hybrid systems where both processes coexist \citep{Silva2021,Mountrichas2023,Byrne-Mamahit2024}. However, in many galaxies, such as those classified as Low-Ionization Nuclear Emission-line Regions (LINERs) or composite systems, the origin of nuclear activity remains ambiguous. Whether the primary energy source is AGN-driven, starburst-driven, or a combination of both remains a topic of debate \citep{Olsson2010,Audibert2021,Zanchettin2024}. Disentangling their contributions is crucial for understanding galaxy evolution, as both AGN feedback and starburst activity regulate the baryon cycle on galactic scales.

The primary challenge stems from the multiwavelength nature of these phenomena, which complicates efforts to resolve degeneracies. A comprehensive approach that integrates radio continuum, molecular line tracers (e.g., CO, HCN, HCO$^+$), infrared diagnostics, and X-ray observations is essential \citep{Herrero-Illana2014,Perez-Torres2021,Radcliffe2021,Farrah2022,Butterworth2025}. Spectral index maps are particularly valuable for distinguishing between thermal free-free emission from HII regions and nonthermal synchrotron radiation originating from supernova remnants or AGN jets \citep{Irwin2019}. In starburst galaxies, galactic-scale winds frequently produce steep-spectrum radio halos in edge-on systems, tracing cosmic-ray electrons in the halo. However, in galaxies with moderate inclinations, disk and halo emissions may overlap spatially, complicating the detection of extraplanar structures \citep{Krause2018}. Furthermore, spatially resolved radio-infrared (IR) correlations can reveal localized deviations from global trends, which can indicate AGN contamination or feedback effects \citep{Cortzen2019,Grundy2023,Vollmer2023}.

In this context, NGC~4527, a nearby edge-on spiral galaxy, serves as an intriguing case study. Classified as a starburst galaxy due to its high infrared luminosity \citep[$L_{\mathrm{IR}}\sim 2.6\times 10^{10}~\mathrm{L_{\odot}}$,][]{Sanders2003} and star formation rate ($\mathrm{SFR}\sim 3 ~\mathrm{M_{\odot}\,yr^{-1}}$), NGC~4527 defies simple classification. Three supernovae have been observed on the galaxy's disk \citep[SN1915A, SN1991T, and SN2004gn,][]{Hakobyan2012}, which provides additional evidence of recent massive star formation activity. 

The nuclear region of NGC~4527 harbours a massive molecular gas reservoir \citep[$M_{\mathrm{H_{2}}}\sim 8\times 10^{9}~\mathrm{M_{\odot}}$,][]{Young1991} and exhibits kinematic signatures of a gravitationally unstable disk ($M_{\mathrm{gas}}/M_{\mathrm{dyn}}=13\%$, \cite{Shibatsuka2003}). However, its star formation efficiency (SFE; $L_{\mathrm{IR}}/M_{\mathrm{H_{2}}}\sim 4.1~\mathrm{L_{\odot}\,M_{\odot}^{-1}}$) is unusually low compared to classic starbursts like M~82 and NGC~253, where $\mathrm{SFE}\gtrsim10$ \citep{Shibatsuka2003}. This discrepancy suggests that NGC~4527 may be in a ``pre-starburst'' phase, where gas accumulation precedes an imminent star formation burst. The main properties of NGC~4527 are listed in Table~\ref{tab:intro}.

NGC~4527 is classified as a LINER, a category often associated with low-luminosity AGNs (LLAGNs). The galaxy hosts an $\mathrm{H_{2}O}$ kilomaser \citep[$L_{\mathrm{H_{2}O}}\sim 4~\mathrm{L_{\odot}}$,][]{Braatz2008}, a feature typically linked to AGN accretion disks or shocked regions in star-forming environments. While early radio studies found no compact flat-spectrum core \citep{Filho2004}, recent X-ray observations have detected a faint nuclear source \citep[$L_{\mathrm{X}}\sim 10^{39.4}~\mathrm{erg\,s^{-1}}$,][]{Hou2024}\footnote{Derived for a distance of $16.5~\mathrm{Mpc}$, equivalent to $L_{\mathrm{X}}=10^{39.2}~\mathrm{erg\,s^{-1}}$ at our adopted distance of 13.5~Mpc.}, providing the first direct evidence of AGN activity. However, we note that the SRG/eROSITA data used for this detection have a spatial resolution of $\sim30''$, making it difficult to determine the origin of the X-ray emission within the galaxy.

High-resolution CO (J = 1-0) observations by \cite{Shibatsuka2003} revealed a strong concentration of molecular gas in the nuclear region ($r<460$~pc) resolved into three distinct components, aligned along the major axis of the galaxy (P.A.=$65^{\circ}$). The central component coincides with the galactic centre and the peak of radio continuum emission, while the two secondary components are symmetrically located at $4''$ from the nucleus, corresponding to peaks in the $10.8~\mu \mathrm{m}$ and radio continuum. Along with these features, two offset ridges of CO emission extend up to $\sim1.8$~kpc, suggesting the presence of a stellar bar. Kinematic analysis reveals a rigidly rotating gas disk with a radius $\lesssim 65$~pc, as well as a CO ring or disk at a radius of 460~pc. This discovery underscores the complexity of LINERs like NGC~4527, where AGN and starburst activity may coexist.

This paper presents new radio continuum observations of NGC~4527 using the upgraded Giant Metrewave Radio Telescope (uGMRT) at 700~MHz (Band-4) and 1230~MHz (Band-5), complemented by archival infrared and X-ray data. Using a multi-wavelength approach, we aim to investigate the enigmatic nature of NGC~4527’s nuclear activity by disentangling the contributions of starburst and AGN processes. The observations and data reduction procedures are detailed in Section~\ref{sec:obs}, where we describe the uGMRT observations and imaging techniques.

Section~\ref{sec:results} presents our findings, including radio continuum morphology, spectral index distribution, and spatially resolved radio–infrared correlation. Finally, Section~\ref{sec:discussion} examines the implications of these results for understanding the nature of nuclear activity in NGC~4527, and Section~\ref{sec:conclusions} summarizes our main conclusions.

\begin{center}
    \begin{table}[t]%
        \centering
        \caption{Properties of NGC~4527.
        \label{tab:intro}}%
        \tabcolsep=0pt%
        \begin{tabular*}{20pc}{@{\extracolsep\fill}lcc@{\extracolsep\fill}}
            \toprule
            \textbf{Parameter} & \textbf{Value} &  \textbf{Reference} \\
            \midrule
            Morphology & SAB(s)bc & (1) \\ 
            RA (J2000) & $12^{\mathrm{h}}34^{\mathrm{m}}08.47''$ & (1) \\
            DEC (J2000) & $02^{\circ}39'14.41''$ & (1) \\
            Distance & 13.5 Mpc & (2) \\
            Optical size & $6.03'\times 1.86'$ & (3) \\
            Inclination & $77^{\circ}$ & (4) \\
            $\log(L_{\mathrm{IR}}/L_{\odot})$ & 10.3 & (5) \\
            SFR & $3~\mathrm{M_{\odot}\,yr^{-1}}$ & (6)\\
            $L_{\mathrm{IR}}/M_{\mathrm{H_{2}}}$ (SFE) & $4.1~\mathrm{L_{\odot}\,M_{\odot}^{-1}}$& (7) \\
            \bottomrule
        \end{tabular*}
        \begin{tablenotes}
            \item[(1)]\; NASA Extragalactic Data Base.
            \item[(2)]\; \cite{Tully1988}.
            \item[(3)]\; SIMBAD Astronomical Database.
            \item[(4)]\; INCLINET \citep{Kourkchi2020}.
            \item[(5)]\; $8-1000~\mu m$ IR luminosity calculated from de IRAS flux following \cite{Sanders1996}.
            \item[(6)]\; $\mathrm{SFR~\left[ M_{\odot}\,yr^{-1}\right]}=0.39\times L_{\mathrm{IR}}/(10^{43}~\mathrm{erg\,s^{-1}})$ \citep{Heesen2018}.
            \item[(7)]\;\cite{Young1991}.
        \end{tablenotes}
    \end{table}
\end{center}

\section{Observations and Data Processing}\label{sec:obs}

The observations of NGC~4527 were conducted using the upgraded Giant Metrewave Radio Telescope~(uGMRT) at 700~MHz (Band-4) and 1230~MHz (Band-5) in May 2022 (proposal ID~42\_016, PI C.A. Galante), with a total on-source time of 1 hour and 45 minutes per band. The flux calibrators, 3C~147 for Band-4 and 3C~286 for Band-5, were observed at the beginning and end of each session, while J~1254+116 served as the phase calibrator, observed between target scans. Detailed observational parameters are provided in Table~\ref{tab:obs}.

Data flagging and calibration were performed using Common Astronomy Software Applications \citep[CASA;][]{CASA2022}. Standard interferometric calibration procedures were applied, including flux, bandpass, and gain calibrations. Self-calibration was applied both in the amplitude and in the phase to improve image quality. For further analysis, visualization, and multiwavelength imaging, we extensively used the {\sc miriad} software package \citep{Sault1995}, {\sc kvis} (from the {\sc karma} package; \citealt{Gooch1996}), and the Cube Analysis and Rendering Tool for Astronomy \citep[CARTA;][]{Comrie2021}. Primary beam correction was performed using the PBCOR task in the Astronomical Imaging Processing System \citep[AIPS;][]{Greisen2003}. 

Imaging was carried out using the CLEAN algorithm with wide field techniques. Natural weighting was applied with a $uv$-range up to 20~k$\lambda$, while uniform weighting used a $uv$-range up to 60~k$\lambda$. The synthesized beams for the naturally weighted images were $16'' \times 13.6''$ at Band-4 and $21.6'' \times 15''$ at Band-5, with RMS noise levels of $0.18~\mathrm{mJy\,beam^{-1}}$ and $0.15~\mathrm{mJy\,beam^{-1}}$, respectively. For uniformly weighted images, the synthesized beams were $4.3'' \times 2.9''$ at Band-4 and $2.5'' \times 2.1''$ at Band-5, with RMS noise levels of $0.15~\mathrm{mJy\,beam^{-1}}$ and $0.06~\mathrm{mJy\,beam^{-1}}$, respectively.  Detailed image parameters are listed in Table~\ref{tab:images}.

\begin{center}
    \begin{table}[t]%
        \centering
        \caption{uGMRT observing parameters.
        \label{tab:obs}}%
        \tabcolsep=0pt%
        \begin{tabular*}{20pc}{@{\extracolsep\fill}lcc@{\extracolsep\fill}}
            \toprule
            & \textbf{Band-4}  & \textbf{Band-5} \\
            \midrule
            Observing date & 02 May 2022 & 06 May 2022 \\  
            Primary beam [arcmin] & 38 & 23 \\
            Time on source [h] & 1.75 & 1.75 \\
            Number of antennas & 29 & 29 \\
            Central frequency [MHz] & 700 & 1230 \\
            Bandwidth [MHz] & 400 & 400 \\
            Flux calibrator & 3C 147 & 3C 286 \\
            Phase calibrator & J1254+116 & J1254+116 \\
            \bottomrule
        \end{tabular*}
    \end{table}
\end{center}

\begin{center}
    \begin{table*}[t]%
        \caption{RMS noise and angular resolution for naturally and uniformly weighted images in Bands 4 and 5.
        \label{tab:images}}
        \centering
        \begin{tabular*}{500pt}{@{\extracolsep\fill}lcccc@{\extracolsep\fill}}
            \toprule
            &\multicolumn{2}{@{}c@{}}{\textbf{Band-4}} & \multicolumn{2}{@{}c@{}}{\textbf{Band-5}} \\
            \cmidrule{2-3}\cmidrule{4-5}
            & \textbf{Natural} & \textbf{Uniform} & \textbf{Natural} & \textbf{Uniform}  \\
            \midrule
            $uv$-range [k$\lambda$] & 0-20 & 0-60 & 0-20 & 0-60 \\
            Resolution $\left[\mathrm{arcsec^{2}}\right]$ & $16\times 13.6$ & $4.3 \times 2.9$ & $21.6\times 15$ & $2.5 \times 2.1$ \\
            Position angle [deg] & 105.2 & 91.1 & 144.7 & 95.9 \\
            RMS noise $\left[\mathrm{mJy\,beam^{-1}}\right]$ & 0.18 & 0.15 & 0.15 & 0.06 \\
            \bottomrule
        \end{tabular*}
    \end{table*}
\end{center}

\section{Results}\label{sec:results}

\subsection{Radio continuum emission}\label{subsec:continuum}

\begin{figure}
    \centering
    \includegraphics[width=8.7cm]{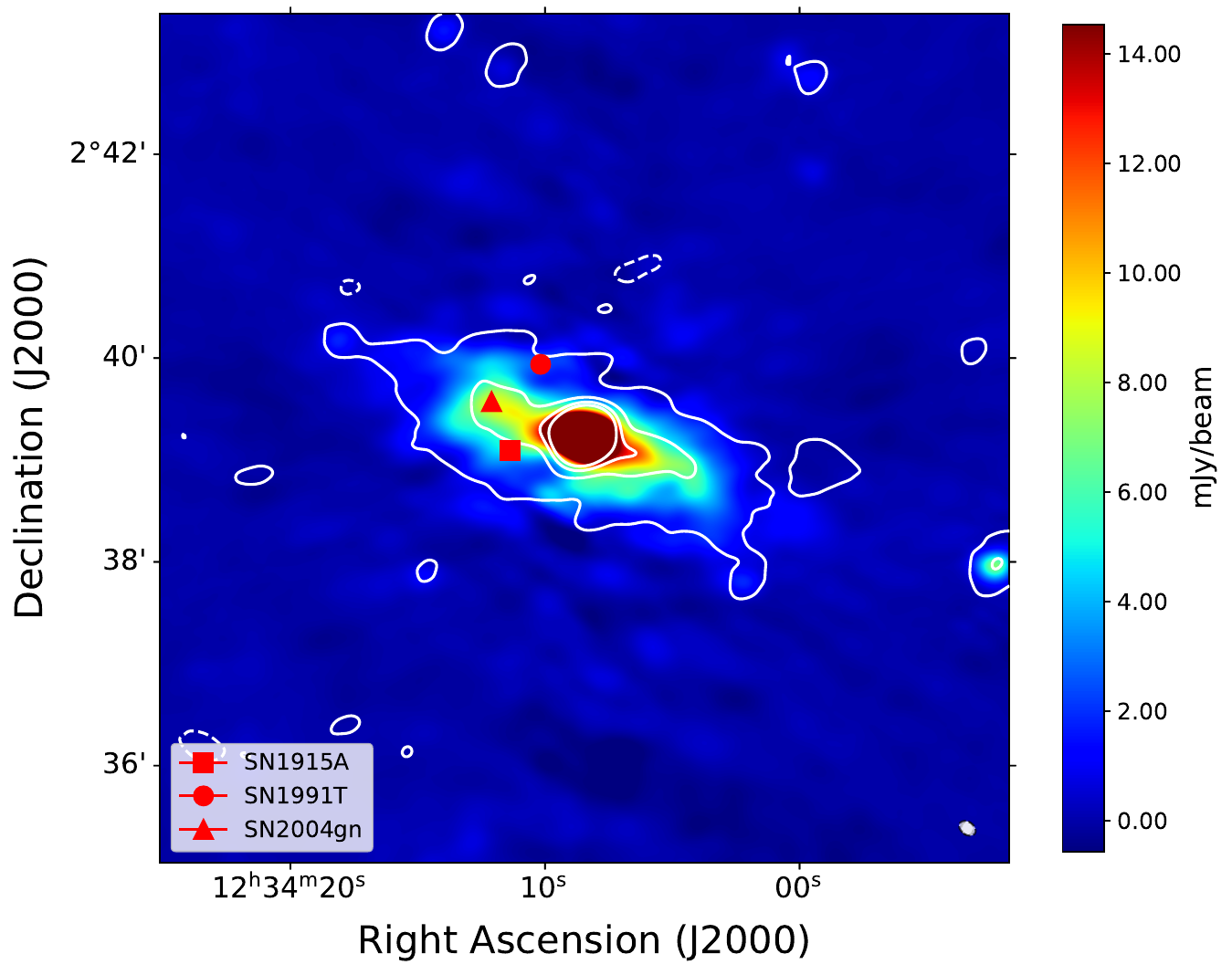}
    \caption{Colour scale shows 700~MHz (Band-4) emission from the naturally weighted image. White contours represent the 1230~MHz (Band-5) emission from the naturally weighted image, with contour levels of 0.45, 4.3 and 8.2~$\mathrm{mJy\,beam^{-1}}$. Red markers indicate the positions of known supernovae in NGC~4527. The synthesized beams are shown in the bottom right corner.}
    \label{fig:cont-nat}
\end{figure}

\begin{figure*}
    \centering
    \includegraphics[width=17.5cm]{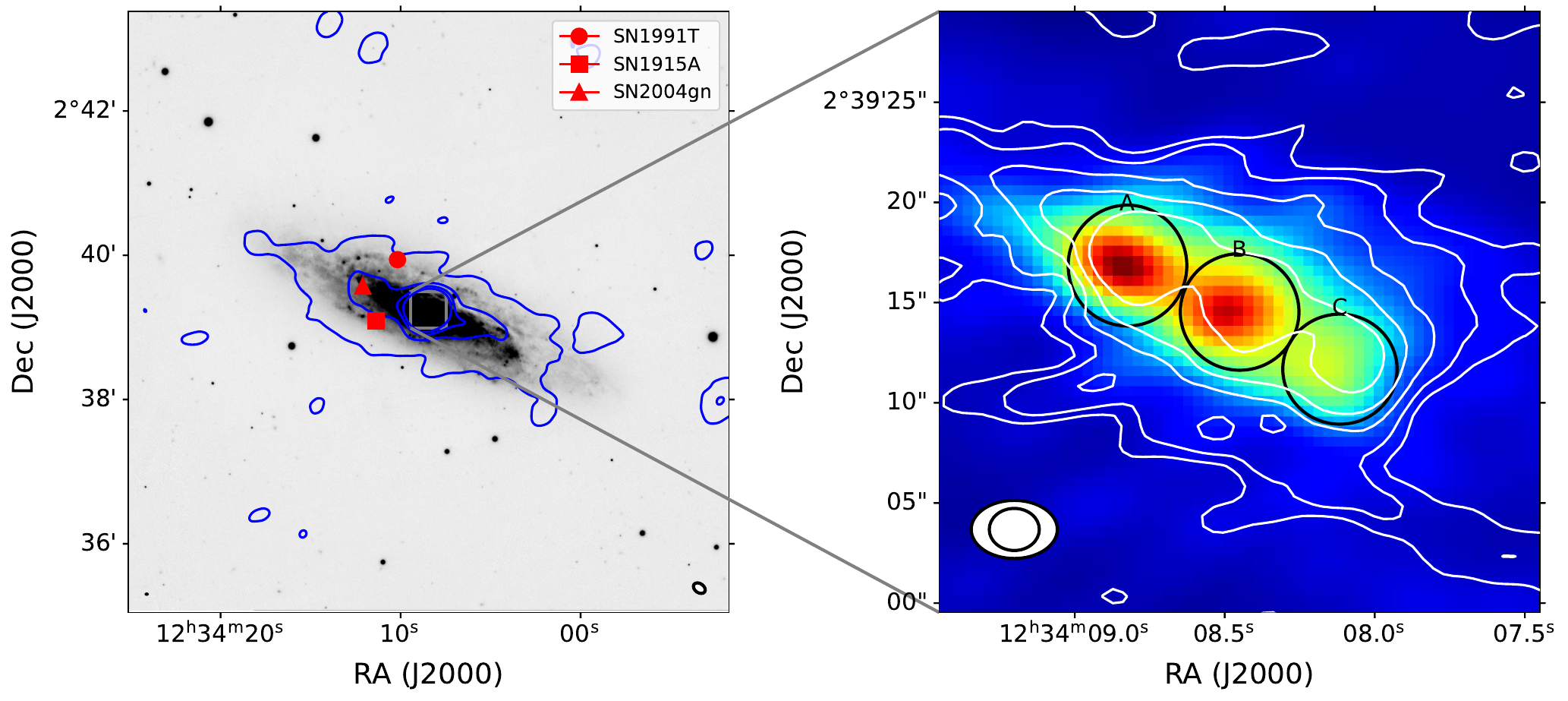}
    \caption{{\bf Left panel:} Blue contours show the 1230~MHz (Band-5) emission from the naturally weighted image, overlaid on the Pan-STARRS $r$-band optical image of NGC~4527. Contour levels are 0.45, 4.3 and 8.2~$\mathrm{mJy\,beam^{-1}}$. Red markers indicate the positions of known supernovae. {\bf Right panel:} Zoom-in view of the central $\sim 30~\mathrm{arcsec^{2}}$ region. Colour scale displays the uniformly weighted 700~MHz (Band-4) image. Overlaid contours correspond to the uniformly weighted 1230~MHz (Band-5) image, showing emission above $3\sigma$ ($\sigma=0.06~\mathrm{mJy\,beam^{-1}}$). The three circular regions marked with labels A, B and C indicate the apertures where flux densities were measured for each source. Synthesized beams are shown in the bottom right corner of both panels.\label{fig:cont-uni}}
\end{figure*}

We present continuum maps of NGC~4527 at Band-4 and Band-5. Low-resolution images were produced using natural weighting and a $uv$-range of 20~k$\lambda$ to maximize sensitivity to diffuse emission. In Fig.~\ref{fig:cont-nat}, Band-4 continuum map is shown in colour scale, with overlaid contours from the Band-5 image for direct comparison. For visual reference of the galaxy's stellar structure, the left panel of Fig~\ref{fig:cont-uni} shows the Band-5 radio contours (same as in Fig.~\ref{fig:cont-nat}) overlaid on a Pan-STARRS $r$-band background image.

The continuum emission in both bands is dominated by a bright central region (hereafter, the "central source") embedded in more extended emission that traces the galaxy's stellar disk. Despite the higher RMS noise in the Band-4 map, its emission appears more extended than that of Band-5. This suggests that the increased spatial extent is intrinsic to the source rather than an artifact of differing map sensitivities.

\begin{figure}
    \centerline{\includegraphics[width=9cm]{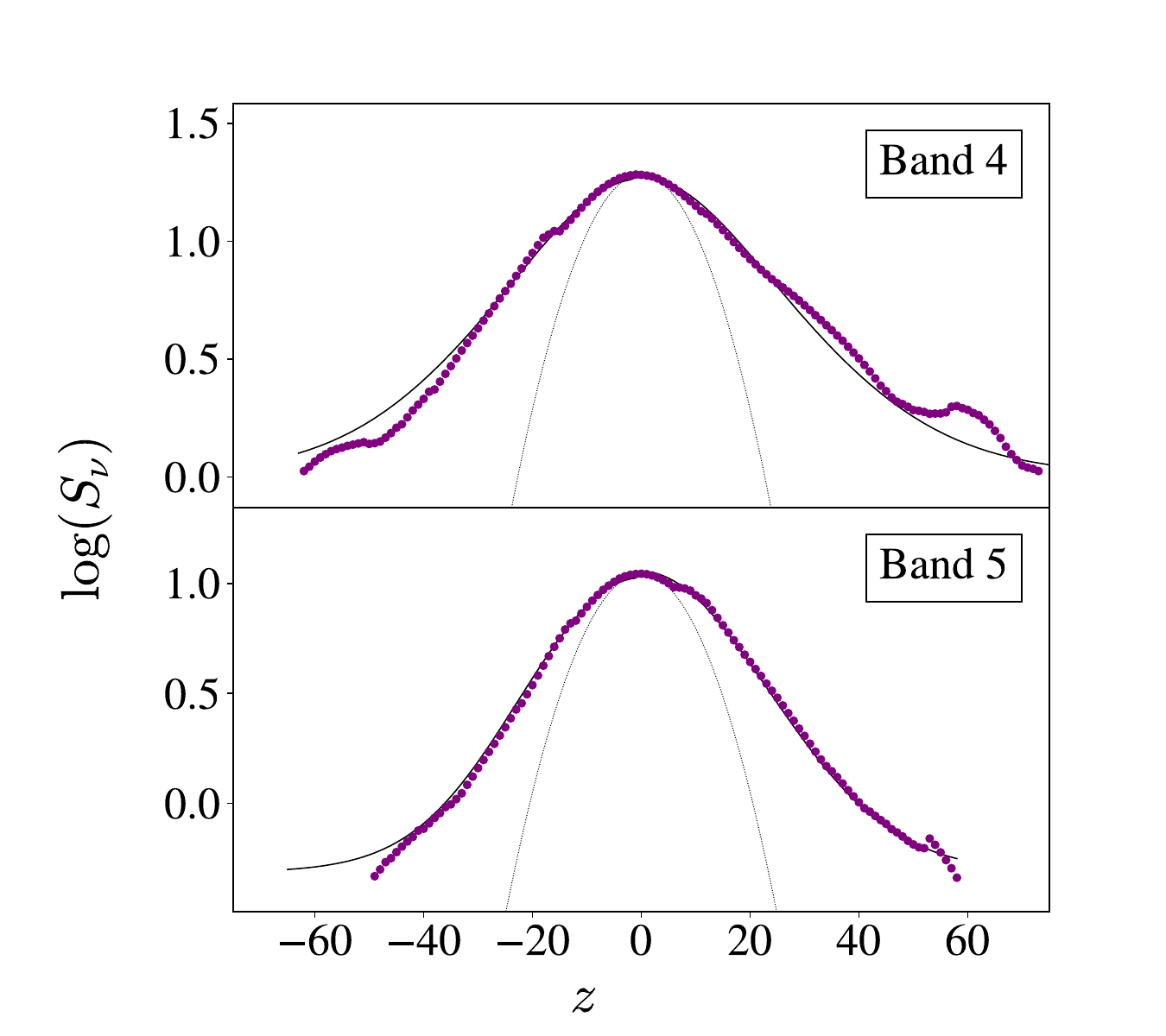}}
    \caption{Average intensity profiles of NGC~4527 in Band-4 (top panel) and Band-5 (bottom panel). The horizontal axis ($z$) represents the vertical offset from the galactic plane ($z=0$) in arcseconds, while the vertical axis shows intensity in units of $\mathrm{mJy,beam^{-1}}$. Scatter points correspond to observed intensity values, and solid lines represent the best-fit one-component exponential models. The synthesized beam is indicated by a dashed line for reference.\label{fig:profiles}}
\end{figure}

The vertical scale height of the radio emission was estimated using the method outlined by \cite{Galante2024}. We fitted a one-component exponential function to the average intensity profiles in both bands, allowing for independent scale heights on either side of the profile while fixing a common central peak. The resulting scale heights from each side were then averaged to obtain a single representative value. The measurements indicate a larger scale height in Band-4 ($13''$ or 0.85~kpc) compared to Band-5 ($9''$ or 0.6~kpc), consistent with the observed maps. The relatively large scale heights in both bands likely reflect the fact that the galaxy is not viewed exactly edge-on. If extraplanar radio emission is present, it would largely overlap with the disk emission, requiring several kiloparsecs of vertical extent above the galactic plane to be distinctly detectable. Figure~\ref{fig:profiles} presents the intensity profiles and their corresponding fitted functions for both bands.

We obtained integrated flux densities for the galaxy by summing all emission above the $3 \sigma$ contour level in the naturally weighted images. The resulting flux densities are $0.43\pm 0.02~\mathrm{Jy}$ at Band-4 and $0.20\pm 0.01~\mathrm{Jy}$ at Band-5. For the central source, we calculated flux densities by summing all emission within a circular region of $20''$ radius centred on the galactic centre. The corresponding flux densities are $0.168\pm 0.008~\mathrm{Jy}$ at Band-4 and $0.094\pm 0.0005~\mathrm{Jy}$ at Band-5, accounting for approximately 40\% and 50\% of the total flux in each band, respectively (See Table~\ref{tab:results-nat} for details). Given the low spatial resolution of these maps ($\sim$1–1.4~kpc), these flux measurements for the central source also include emission from the circumnuclear region. 

To mitigate the contribution of circumnuclear extended emission to the central source, we produced high-resolution images using uniform weighting and extended baselines, as shown in the right panel of Fig.\ref{fig:cont-uni}. In these maps, the emission is confined to the central $\sim 30''$ of the galaxy, and the extended disk is not recovered. In addition, the central source is resolved into three distinct components: source B, located at the centre, and sources A and C, symmetrically located along the major axis of the galaxy at a distance of approximately $6''$ (400~pc) from the centre. The flux densities of source B are $23 \pm 1~\mathrm{mJy}$ in Band-4 and $17.1 \pm 0.9~\mathrm{mJy}$ in Band-5. For source A, the flux densities are $23 \pm 1~\mathrm{mJy}$ in Band-4 and $17.1 \pm 0.1~\mathrm{mJy}$ in Band-5. Source C exhibits flux densities of $13.3 \pm 0.9~\mathrm{mJy}$ in Band-4 and $11.2 \pm 0.6~\mathrm{mJy}$ in Band-5. These values were obtained by summing the emission within non-overlapping circular regions of $\sim3''$ radius, each centred on the corresponding source. A summary of all measurements is provided in Table~\ref{tab:results}.

\subsection{Spectral index distribution}\label{subsec:specindex}

To examine the spatial distribution of the spectral index\footnote{In this work, we adopt the convention $S_{\nu} \propto \nu^{\alpha}$, where $S_{\nu}$ is the flux density at frequency $\nu$, and $\alpha$ is the spectral index.} across NGC~4527, we generated a spectral index map between 700~MHz (Band-4) and 1230~MHz (Band-5) using naturally weighted images. Both images were created using a common $uv$-range of 20~k$\lambda$, ensuring that comparable spatial scales are sampled in both bands. To achieve spatial consistency, the images were regridded and convolved with a $22''$ Gaussian beam, enabling a one-to-one correspondence across the bands. The resulting map includes only regions where the emission exceeds three times the noise level in both bands.

\begin{figure*}
    \centerline{\includegraphics[width=18cm]{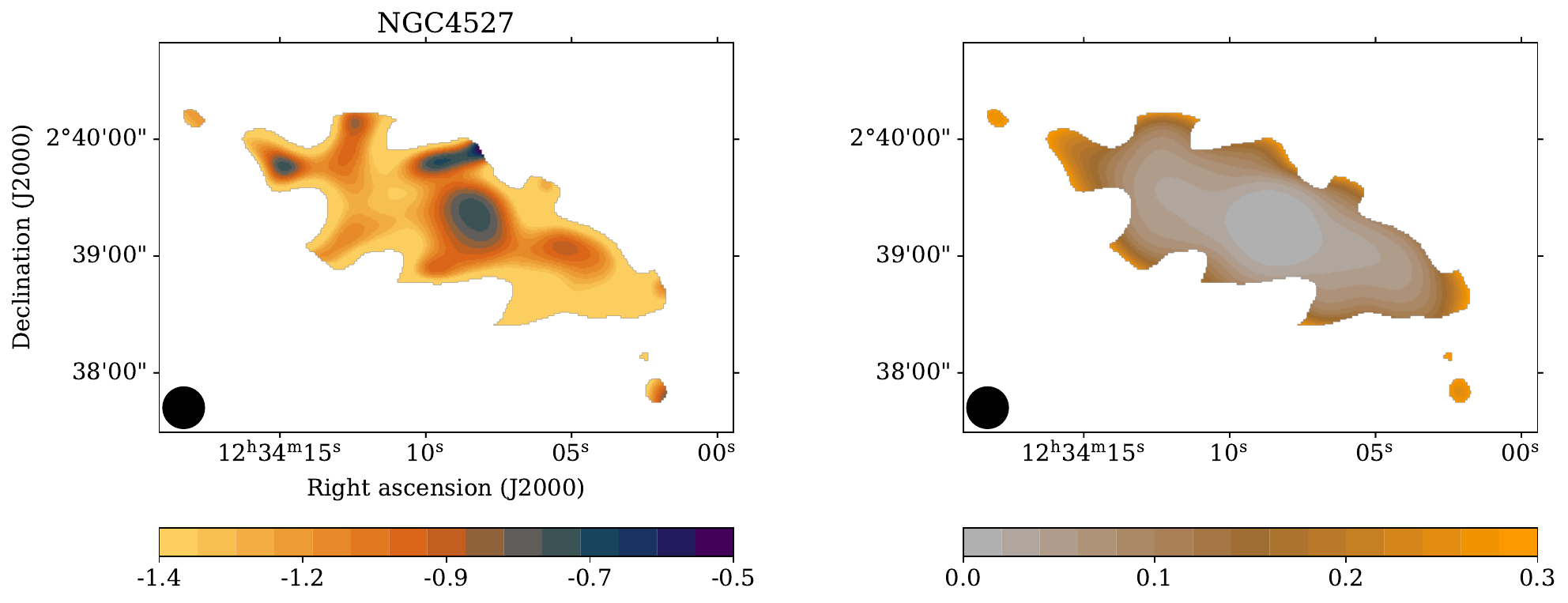}}
    \caption{Spectral index map (left) and error map (right) of NGC~4527, based on 700~MHz and 1230~MHz images. Spectral indices were computed for pixels with intensities above three times the noise in both bands. Flatter indices are shown in dark blue, while steeper indices appear in orange and yellow \citep{English2024}. The beam size is indicated in the bottom-left corner of each panel.}\label{fig:spectral_index}
\end{figure*}

The resulting spectral index map is shown in Fig.~\ref{fig:spectral_index} alongside its corresponding error map. Morphologically, the central region and its immediate surroundings exhibit the flattest spectral indices, around $-0.8$, while the rest of the galactic disk displays steeper values, typically near or below $-1$. Spectral indices tend to become steeper with increasing distance from the galactic plane, with the exception of flat values near the map edges that are likely artifacts resulting from higher uncertainties rather than physical features. Overall, the errors are smallest in the central region, progressively increasing toward the map’s edges, ranging from 0.01 to 0.25, which corresponds to approximately 2\% to 20\% of the measured values.

Spectral index calculations for the isolated central sources, based on uniformly weighted images that minimize extended emission and isolate compact sources, yielded values of $-0.3\pm 0.1$ for source A, $-0.5\pm 0.1$ for source B, and $-0.3 \pm 0.1$ for source C. These results are summarized in Table~\ref{tab:results}.

\begin{center}
    \begin{table}[t]%
        \centering
        \caption{Physical parameters of NGC~4527 derived from uGMRT naturally weighted images at 700~MHz (Band-4) and 1230~MHz (Band-5), separated into galaxy-scale properties and central source characteristics.}
        \label{tab:results-nat}
        \tabcolsep=0pt%
        \begin{tabular*}{20pc}{@{\extracolsep\fill}lcc@{\extracolsep\fill}}
            \toprule
            \textbf{Parameter}                                  & \textbf{Band-4}  & \textbf{Band-5} \\
            \midrule
            \multicolumn{3}{@{}l}{\textbf{Galaxy}} \\
            $S_{\nu}$ (mJy)                                     & $430 \pm 2$      & $200 \pm 1$ \\
            $L_{\nu}$ ($10^{36}$ erg s$^{-1}$)                  & 23.45            & 15.27 \\
            $\langle z \rangle$ (kpc)                           & $0.85 \pm 0.01$  & $0.59 \pm 0.01$ \\

            \multicolumn{3}{@{}l}{\textbf{Central source}} \\
            $S_{\nu}$ (mJy)                                     & $168 \pm 8$      & $94 \pm 5$ \\
            $L_{\nu}$ ($10^{36}$ erg s$^{-1}$)                  & 9.16             & 7.18 \\
            $S_{\mathrm{peak}}$ (mJy beam$^{-1}$)               & 94.75            & 65.08 \\
            \bottomrule
        \end{tabular*}
        \begin{tablenotes}
            \item \footnotesize Note: $S_\nu$ is the integrated flux density. For the galaxy, it was calculated by summing all emission above the $3\sigma$ contour in each image. For the central source, it was measured within a circular region of $20''$ radius centred at the galaxy's coordinates (See Table~\ref{tab:intro}). $S_{\mathrm{peak}}$ is the maximum pixel value within that region. $L_\nu$ is the corresponding radio luminosity, and $\langle z \rangle$ is the mean vertical scale height.
        \end{tablenotes}
    \end{table}
\end{center}

\begin{center}
    \begin{table*}[t]%
        \centering
        \caption{Properties of the nuclear sources identified in NGC~4527 using uniformly weighted images from uGMRT Bands 4 and 5.}
        \label{tab:results}
        \tabcolsep=0pt%
        \begin{tabular*}{400pt}{@{\extracolsep\fill}lccccccc@{\extracolsep\fill}}
            \toprule
            \textbf{Source} & \textbf{Band} & $\mathbf{S_{\nu}}$ & $\mathbf{L_{\nu}}$ & $\mathbf{S_{\mathrm{peak}}}$ & $\mathbf{\Delta \alpha}$ & $\mathbf{\Delta \delta}$ & $\mathbf{\alpha}$ \\
            & & (mJy) & ($10^{36}~\mathrm{erg\,s^{-1}}$) & $\left(\mathrm{mJy\,beam^{-1}}\right)$ & (s) & ( $''$ ) & \\
            \midrule
            \textbf{A} & 4 & $23 \pm 1$     & 1.25 & 17.3 & \multirow{2}{*}{0.36}  & \multirow{2}{*}{2.44}  & \multirow{2}{*}{$-0.3 \pm 0.1$} \\
           & 5 & $19 \pm 1$     & 1.45 & 9.2  &  &  &  \\
            \textbf{B} & 4 & $23 \pm 1$     & 1.25 & 16.3 & \multirow{2}{*}{-0.01} & \multirow{2}{*}{-0.21} & \multirow{2}{*}{$-0.5 \pm 0.1$} \\
           & 5 & $17.1 \pm 0.1$ & 1.31 & 6.9  &  &  &  \\
            \textbf{C} & 4 & $13.3 \pm 0.8$ & 0.73 & 10.3 & \multirow{2}{*}{-0.35} & \multirow{2}{*}{-2.73} & \multirow{2}{*}{$-0.3 \pm 0.1$} \\
           & 5 & $11.2 \pm 0.6$ & 0.85 & 4.4  &  &  &  \\
            \bottomrule
        \end{tabular*}
        \begin{tablenotes}
            \item \footnotesize Note: The integrated flux density ($S_\nu$) and the peak intensity ($S_{\mathrm{peak}}$) were measured within circular regions of $3''$ radius, centred on the positions given by the right ascension and declination offsets ($\Delta \alpha$, $\Delta \delta$), which are measured relative to the galaxy's coordinates (See Table~\ref{tab:intro}). The radio luminosity ($L_\nu$) was computed from the integrated flux. The spectral index ($\alpha$) was derived between Bands 4 and 5.
        \end{tablenotes}
    \end{table*}
\end{center}

\subsection{Radio-IR relation}

The correlation between radio and infrared (IR) emission in galaxies, known as the radio–IR relation, has been extensively studied, particularly in star-forming systems. This relation connects the radio continuum luminosity—primarily tracing synchrotron emission from cosmic-ray electrons accelerated by supernovae—with the IR luminosity, which predominantly arises from thermal dust emission heated by young, massive stars. Remarkably, the correlation spans several orders of magnitude in both radio and IR luminosities, making it a powerful diagnostic tool in extragalactic astronomy \citep[e.g.,][]{Helou1985, Condon1992, Delhaize2017, Delvecchio2021, Molnar2021}. It has been rigorously tested at low redshifts, where its robustness has been confirmed among various types of galaxies, including merging systems \citep{Condon2002, Murphy2013}.

The radio-IR relation is often expressed in terms of the parameter $q$, defined as:

\begin{equation}
    q=\log \left(\frac{L_{\mathrm{IR}}}{3.75\times10^{12}\,\mathrm{W}}\right)-\log \left(\frac{L_{1.4}}{\mathrm{W\,Hz^{-1}}}\right),
\end{equation}
\\
\noindent where $L_{\mathrm{IR}}$ is the IR luminosity and $L_{\mathrm{1.4}}$ is the radio continuum luminosity at $1.4~\mathrm{GHz}$. The value of $q$ reflects the balance between star formation, which drives both IR and radio emission, and additional processes such as AGN activity or environmental effects that can disrupt the correlation. Although the precise origin of the radio-IR relation remains an open question, it is commonly attributed to the interplay of physical mechanisms governing star formation.

In \cite{Grundy2023}, the authors explored a modification of the radio–IR relation by using the Wide-field Infrared Survey Explorer (WISE) W3 band in place of the total or far-infrared luminosity, as is commonly done in most analyses. Among the bands observed by WISE, the W3 band serves as an excellent tracer of interstellar medium (ISM) emission. They demonstrated that the resulting $q_{\mathrm{W3PAH}}$ parameter provides a useful diagnostic for distinguishing between AGN activity and star formation. This band includes significant contributions from polycyclic aromatic hydrocarbons (PAHs), particularly the prominent emission feature at $11.3~\mu \mathrm{m}$. The flux in the W3 band associated with PAHs (W3PAH) is widely used as a tracer of the star formation rate (SFR), offering an alternative to the total IR or far-IR fluxes traditionally employed. A major advantage of using W3PAH lies in the high angular resolution of WISE observations, which enables spatially resolved analyses of the radio–IR relation across a galaxy. PAH emission is associated with diffuse radiation fields in the ISM produced by star-forming regions; however, PAHs can also be destroyed by intense radiation fields generated by AGNs and massive stars, as well as by shocks from supernovae, galaxy interactions, or mergers.

\begin{figure}
    \centerline{\includegraphics[width=8cm]{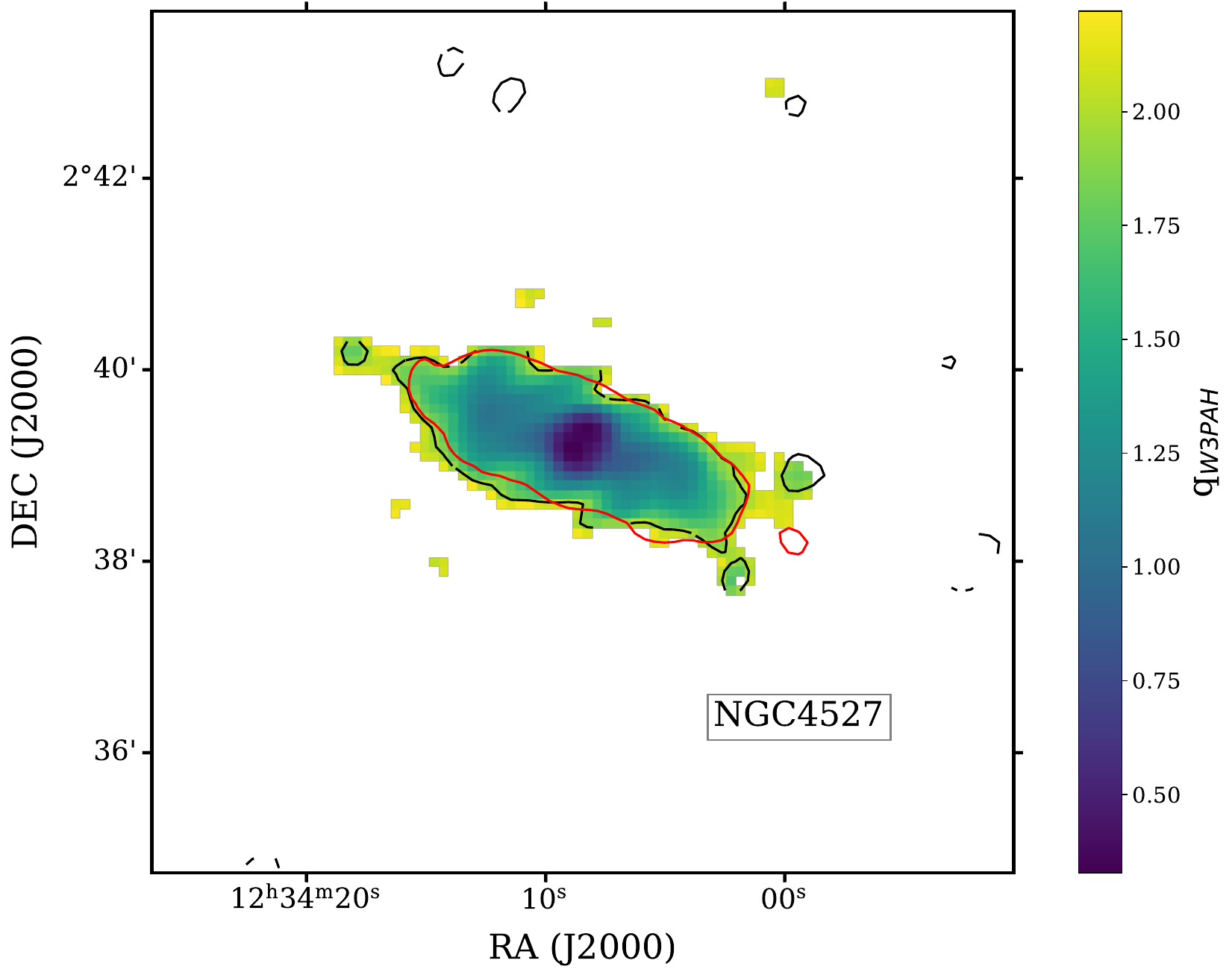}}
    \caption{Spatially resolved map of $q_{\rm W3PAH}$ across NGC~4527. Black and red contours indicate regions where radio (uGMRT Band-5) and infrared (WISE W3) emission, respectively, exceed three times the noise level.}\label{fig:qmap}
\end{figure}

For this study, we followed the procedure outlined in \cite{Grundy2023} to obtain the WISE IR luminosity and generated a $q_{\mathrm{W3PAH}}$ map for NGC 4527, shown in Figure~\ref{fig:qmap}. The WISE data, which are publicly available, were retrieved from the NASA/IPAC Infrared Science Archive (IRSA). To compute the radio luminosity, we used the uGMRT Band-5 data. In the figure, black and red contours delineate regions where the emission exceeds three times the noise level in the radio and IR bands, respectively. The values of $q_{\mathrm{W3PAH}}$ were calculated only for pixels where both the radio and IR emission surpassed this threshold, ensuring that the measurements are not biased by regions lacking significant signal in either dataset.

This map reveals significant variation in $q_{\mathrm{W3PAH}}$ across the galaxy. In the main disk, we find typical $q_{\mathrm{W3PAH}}$ values of approximately 1, indicating an environment dominated by diffuse radiation fields associated with moderate star formation activity. In the central regions, where the radiation field intensity is higher, $q_{\mathrm{W3PAH}}$ decreases to around 0.5, suggesting PAH destruction in these more energetic zones. In contrast, in the outer regions of the galaxy, $q_{\mathrm{W3PAH}}$ increases again to $\gtrsim 1.75$, reflecting conditions more favourable for PAH preservation. These results are in agreement with previous findings. \cite{Grundy2023} reported $q_{\mathrm{W3PAH}}$ values ranging from 1.2 to 1.65 in normal star-forming galaxies, and around 0.6 in the central region of NGC~1367, which harbour an AGN. Similarly, \cite{saponara2025} reported $q_{\mathrm{W3PAH}}$ values of approximately 0.5 in NGC~6221 and NGC~3256, which they suggest host LLAGNs, while non--active galaxies in their sample showed values between 1 and 3.

\section{Discussion}\label{sec:discussion}

The radio–IR correlation presented in Fig.~\ref{fig:qmap} suggests that the star formation activity is concentrated on the disk. Sources A and C exhibit slightly flatter spectral indices, likely resulting from a mixture of thermal emission from HII regions and nonthermal radiation from supernova remnants and colliding stellar winds. The destruction of PAHs in the central region, which coincides with the galactic centre, might be attributed to X-ray irradiation from the accretion disk around the central black hole. 

Using the tight empirical relation between the black hole mass $M_{\mathrm{BH}}$ and the central velocity dispersion $\sigma_{c}$ \citep{Ferrarese2000, Gebhardt2000, Tremaine2002} and adopting a value of $\sigma_{\mathrm{c}}=214~\mathrm{km\,s^{-1}}$, \cite{Filho2004} inferred a black hole mass of $M_{\mathrm{BH}}\sim 10^8~\mathrm{M_{\odot}}$ and an associated Eddington luminosity $L_{\mathrm{Edd}}\sim 10^{46}~\mathrm{erg\,s^{-1}}$.

The position of NGC~4527 in the diagram of the relation between the milliarcsecond-scale radio core luminosity at 5~GHz and the black hole mass ($L_{\mathrm{R}}\propto m_{\mathrm{acc}}^{1.2}\,M_{\mathrm{BH}}^{1.6}$, see Fig.~7 of \citealt{Filho2004}) might be indicative of a scenario with an advection-dominated accretion flow (ADAF) with an accretion rate of $m_{\mathrm{acc}}\lesssim10^{-4}$. However, a high accretion rate scenario cannot be ruled out, because the galaxy is seen edge-on; in the case of super-Eddington accretion, the dense wind ejected by the critical part of the accretion disk \citep{Sotomayor2022} would absorb most of the central X-ray emission \citep{Zhou2019ApJ, Abaroa2024A&A}. In such a scenario, the interaction of the disk wind with clouds and stars orbiting the black hole could result in radio emission from electrons accelerated at the resulting bow shocks \citep{Sotomayor2022}. This mechanism naturally leads to canonical spectral indices of $\sim 2$ for the electrons and $\sim -0.5$ for the optically thin radio flux, as we report for source B (see \citealt{Muller2020MNRAS, Sotomayor2022}). 

The lack of evidence for either a jet or a radio halo in our high-resolution radio images supports the hypothesis that only a weak wind is being ejected from the accretion disk. Given that the escape velocity for a black hole of mass $\sim 10^8~\mathrm{M_{\odot}}$ is approximately 10\% of the speed of light, a wind launched during a super-Eddington accretion episode would likely fall back to the galactic plane. Such winds typically exhibit a strong equatorial component (e.g., \citealt{Yoshioka2022PASJ}) and would therefore fall upon the plane at large distances from the central source. Assuming a symmetric accretion process, the fallback material would likely settle into a ring around the galactic centre, injecting fresh gas and potentially triggering star formation. Indeed, a molecular gas ring was detected by \cite{Shibatsuka2003}, who identified two sources coinciding with sources A and C and carried out a dynamical analysis of the structure. Similar circumnuclear rings have also been observed in other nearby star-forming galaxies with composite or LINER-like nuclear activity \citep{Olsson2010, Lai2022, Audibert2019, Audibert2021}, suggesting that this configuration may be a common outcome of evolutionary processes in spiral galaxies.

Assuming a super-Eddington transient phase lasting approximately $10^6$yr with an accretion rate ten times the critical value, we estimate that roughly $3.5\times10^6\mathrm{M_{\odot}}$ of gas would have been deposited onto the galactic plane. This inflow of fresh material could have triggered the star formation episodes that are still ongoing at the locations of sources A and C. In contrast, the central accretion episode may have ended some time ago, consistent with the low accretion rate onto the central black hole inferred by \cite{Filho2004}.

The alternation between high- and low-accretion phases, accompanied by episodes of intense star formation, may play a key role in shaping the activity and evolution of NGC~4527. Owing to its proximity, NGC~4527 represents an ideal target for future observations aimed at testing models of how accreting supermassive black holes influence the evolution of their host galaxies through large-scale outflows and multiphase feedback (e.g., \citealt{Kormendy2013ARA&A}).

\section{Conclusions}\label{sec:conclusions}

In this work, we have presented a radio study of the edge-on starburst–LINER galaxy NGC~4527 based on uGMRT observations performed at 700 and 1230~MHz. By combining interferometric imaging with ancillary infrared WISE data, we investigated both the large-scale emission and the central region of the galaxy.

On large scales, the radio emission at both frequencies follows the stellar disk, with no evidence of a radio halo. The spectral index values across the disk are consistent with nonthermal synchrotron emission from star-forming regions, and the spatially resolved radio–infrared correlation --tracing the presence of PAHs-- supports the interpretation of NGC~4527 as a typical star-forming spiral galaxy.

In the nuclear region, we resolved three compact radio sources: one at the galactic center, which exhibits a nonthermal spectral index, and two symmetrically located along the galaxy’s major axis at a projected distance of approximately 400~pc, both displaying relatively flat spectral indices. This configuration is consistent with the presence of a circumnuclear star-forming ring, as previously inferred from CO observations by \cite{Shibatsuka2003}. The combination of PAH destruction, nonthermal radio emission and the recent detection of X-rays in the central region can be explained by the presence of an active galactic nucleus.

Moreover, the AGN scenario offers a plausible mechanism for the formation of the circumnuclear ring: if the accretion flow is --or was-- super-Eddington and driving a dense wind that fails to escape the galaxy’s gravitational potential, the fallback of material onto the disk could inject fresh gas a few hundred parsecs from the centre, triggering star formation in a ring-like structure.

The results presented in this paper highlight the importance of multi-frequency studies in disentangling the origin of radio emission and the physical processes shaping the nuclear regions of galaxies. Confirming the presence of an AGN in NGC~4527 will require variability studies and higher-resolution observations of the innermost central region.

\section*{Acknowledgments}
  G.E.R. acknowledges financial support from the State Agency for Research of the Spanish Ministry of Science and Innovation under grant PID2022-136828NB-C41/AEI/10.13039/501100011033/, and by “ERDF A way of making Europe”, by the “European Union”. He also thanks support from PIP 0554 (CONICET).

\bibliography{Paper-NGC4527-clean}%

\end{document}